# Pristine graphene as a catalyst in reactions with organics containing C=O bonds

Xiaozhi Xu[2,3], Zhiyuan Liu[3], Fen Ke[3,4], Chenfang Lin[3,5], Kaihui Liu[3], Xia Guo[1], Zhaohui Zhang[3], Xinzheng Li*[3], Zonghai Hu*[1]

[1] School of Electronic Engineering, State Key Laboratory for Information Photonics and Optical communications, Beijing University of Posts and Telecommunications, Beijing 100876, China

[2] School of Physics and Telecommunication Engineering, South China Normal University, Guangzhou 510006, China

[3] State Key Laboratory for Mesoscopic Physics, School of Physics, Peking University, Beijing 100871, China

[4] Institute of Physics, Chinese Academy of Sciences, Beijing 100190, China

[5] Department of Chemistry, University of Liverpool, Liverpool, United Kingdom

[#] These authors contributed equally to this work.

*Correspondence and requests for materials should be addressed to: guox@bupt.edu.cn, xzli@pku.edu.cn

**Pristine graphene is thought lack of catalytic activity up to date, although using graphene-plus-heteroatom materials as catalysts has become a subject of intensive research because it can be metal saving, eco-friendly and ultimately sustainable. Here we report observations of catalytic reactions of high-quality, clean, pristine graphene when immersed into organics containing C=O bonds, like acetone, acetic acid and acetaldehyde. The C=O bonds were found to break and form polymers including polyethers. The reaction rate is highly temperature dependent. The reaction products mainly physically adsorb on graphene and do not cause increase of defect density in graphene, hence graphene retains its intrinsic properties. This new catalysis shall not only find practical importance but also deepen our understanding on the role of graphene in all graphene based catalysis.**

Searching for new catalysis has always attract great research interests because about 90% worldwide chemical products involves catalysts in the production process[1]. Traditional metal or metal compound catalysts are highly efficient but their sustainability is questionable due to limited natural resouce[2]. Besides, noble metals are quite expensive and many post-reaction transition metals are toxic, causing enviromental concerns. Therefore metal-free catalysts are highly desirable. Carbon based catalysts have seen great potential in this regard[3-13]. Since the discovery of graphene[14], graphene based catalysts have received more and more attention because of many amazing properties of graphene, for one instance, the huge specific suface area of ~2600 $m^2/g$[15]. Graphene oxide and reduced graphene oxide have found use in many catalysis[16-18]. However, residual metal contents in them are unavoidable thus the true active sites and catalytic mechanisms are still in debate[3]. Graphene with nonmetallic dopants such as N, O, B, P and Si has been demonstrated catalytic in many reactions including oxygen reduction reactions, oxygen evolution reactions and hydrogen evolution reactions[19-23]. Meanwhile, as a logical and basic step to enhance the catalytic effeciencies of these catalysts and find/design new ones, study on catalysis of “pure” (heteroatom-free) graphene, is somehow missing. One possible reason is that the mass yield of “pure” graphene is quite low, either by mechanical exfoliation or chemical vapor deposition[24-27]. Moreover, “pure” graphene has long been considered mostly inert and lack of catalytic activity[28, 29]. On the other hand, when transferring graphene, there are often contaminants left, affecting subsequent device performance. Chemicals such as PMMA can stay on graphene because of the pi-pi stacking interaction. Acetone is widely used as a rinse solvant to remove PMMA and clean graphene surface[30-33]. It is least considered to react with pristine graphene. In this work, we found that chemical reaction of acetone can occur at the presence of high-quality, clean, pristine graphene, forming polymers on the surface. The C=O bonds break and form chain-like and cyclic polymers under the catalytic influences of graphene. The polymers give rise to extra Raman peaks under 633 nm excitation. However, graphene remains intact with no sign of increased defect density. This discovery opens the door of using pristine graphene as a metal-free, dopant-free catalyst and should have important implications on organic chemistry and graphene device fabrication.

**Results**

**Discovery of extra Raman peaks of graphene after immersed into pure acetone.**

The graphene samples were grown by atmospheric pressure chemical vapor deposition (APCVD) method (See Methods for growth details). No D peaks were observed in the Raman spectra of the as-grown graphene samples (Fig. 1a-b), indicating the high quality of graphene with a negligible defect density. To investigate whether actone reacts with graphene and avoid complexity, no chemicals like PMMA were used and the graphene samples were only immersed into pure acetone. After 10 hrs of immersion, Raman spectra with 514 nm excitation show no changes (Fig. 1b). However, if 633 nm excitation wavelength is used, extra peaks other than the graphene G and 2D bands emerge (Fig. 1b and Supplementary Fig. 1). Raman signals of intrinsic graphene would be resonant with all excitation wavelengths because of the Dirac cone band structure. Therefore the extra peaks resonant with 633 nm but not 514 nm excitation should be coming from some molecules other than graphene itself. Raman spectra of pure acetone and pristine graphene (Fig. 1b-c) can exclude the signal coming from residual acetone. The increase of the intensity of the extra Raman peaks with longer immersion time (Fig. 1d) shows the progression of the reaction between acetone and the samples. However, the defect density of graphene remains negligible since the Raman D peak is still absent. Moreover, annealing the samples at 170 °C for 1 hr in air after immersion could not efficiently remove the extra peaks (the molecules) (Fig. 3a).

To check if acetone reacts with the Cu substrates instead of graphene, Raman measurements were performed in bare Cu surface areas not covered by graphene (Fig. 1a). The extra Raman peaks do not appear in bare Cu areas (Fig. 2a). Other substrates including rough ground glass, $SiO_2$, GaN, mica, and layered $MoS_2$ do not show signs of reaction with acetone after immersion(Fig. 2a), either. Freshly cleaved highly oriented pyrolytic graphite (HOPG), after immersion, gives the same extra Raman peaks as monolayer graphene does. These results demonstrate that pristine graphene (metal-free, dopant-free) do react with acetone to give rise to the observed extra Raman peaks.

**Initial reaction: broken of the C=O bond.**

Next, we carried out experiments to pinpoint the exact bond to which the reaction initially happens. For this purpose, various organic solvents besides acetone were used for immersion. After as-grown graphene was immersed into pure acetic acid, the same extra Raman peaks were detected (Fig. 2b). Both solvents have one C=O bond. The only difference between the molecular structure of acetone and acetic acid is that one C-$CH_3$ bond in acetone is substituted by one C-OH bond in acetic acid (Fig. 2b). Yet these two solvents result in the same Raman peaks. This comparison study indicates that the initial reaction happens to the C=O bond, not the C-$CH_3$ bonds or the C-OH bond. If as-grown graphene was immersed into pure ethanol, the above mentioned extra Raman peaks do not appear at all. The only structural difference here is that the C=O bond in acetic acid is substituted by two C-H bonds in ethonal (Fig. 2b). Acetic acid causes the extra Raman peaks while ethonal does not. This further supports the claim that the initial reaction happens to the C=O bond. Following this reasoning with ketone and carboxylic acid, we turn to another type of organics that possesses the C=O bond, the aldehyde. Indeed, the Raman peaks again emerge after ethanal immersion. These results demonstrate that the C=O bond is the reactive bond with pristine graphene.

**Physical adsorption of the product molecules.**

After the C=O bond is broken, two dangling bonds are left. One on the C end and the other on O. It appears unlikely that they simply bond to graphene. Because the intensity of the extra Raman peaks indicates a sufficient amount of product molecules. Have the dangling bonds chemically bond to graphene, it would cause a significant increase of defect density in graphene thus giving rise to a detectable Raman D peak[34], which was not the case. If most of the dangling bonds do not form new bonds with graphene, then they should form bonds with other broken C=O bonds to lower the energy and stabilize. Depending on which end meets, this can result in sequences like -C-O-C-O-, -O-C-C-O- and -C-O-O-C-. Since -C-O-O-C- (peroxide) is usually much less stable, the first two sequences should dominate. The two sequences can mix up to form chains. Most of these product chains should not be too short. Otherwise they either chemically adsorb on graphene to cause a Raman D peak, or physically (weakly) adsorb. But annealing the samples at 170 °C was not efficient for their desorption (cleaning) (Fig. 3a). Longer chains have large number of carbon atoms. Even the van der Waals interaction per atom between the chain and graphene is quite small, the total interaction is significant, consistent with the apparent high desorption temperature (Supplementary Fig. 2). The characteristic structural unit in the long chains is –C-O-C-, the ether linkage. So the main products of the graphene/C=O reaction should include polyethers, and probably polyketals- and/or polyacetals. The product polyethers can take various forms including open-ended or cyclic. Scanning tunneling microscopy (STM) images (Fig. 3b-d and Supplementary Fig. 3) taken after acetone immersion and mild annealing in UHV confirmed the existence of product molecules of long chain form and cyclic form. High resolution image of the cyclic molecule marked by “A” shows the structure that mimics the 18-crown-6 crown ether[35]. We note that during tip scanning, the chain-like and cyclic features tend to move and deform. This again implies the weak, physical adsorption nature rather than strong, chemical adsorption.

**DFT calculations and reaction mechanism.**

The above systematic experimental studies show that when pristine graphene is immersed into organic solvents, it can make the double C=O bonds of the solvent molecules break and form new polymers. To check the energetics of this picture, we used a simple model consisting of two acetone molecules and a graphene monolayer with one missing C atom and performed density functional theory (DFT) simulations. The structure of the acetone molecules were relaxed in the gas phase and the geometry optimized structure of the acetone molecules adsorbing on graphene is shown in Fig. 4a-b. It is clear that the double C=O bonds break and one C-O-C ether group links both acetone molecules. In addition, the total energy of this shown state is 0.502 eV lower than that of the initial state consisting of two free acetone molecules and a free graphene monolayer. This is consistent with our experimental results. We note that though defect sites may make the reaction easier to happen, the reaction products are not confined to the defect sites or limited by the defect density. In the middle of large single-crystal graphene grains (sub-millimeter in diameter) where defects were almost nonexistent (as shown by the negligible Raman D peak and STM images), a significant amount of reaction products were still detected by Raman with no smaller intensity. Also on freshly cleaved HOPG where pre-adsorbed species were very few, Raman signals of the reaction products were detected. Therefore the role of pristine graphene could well be just lowering the reaction barrier, i.e. the role of a catalyst. Figure 4c shows the temperature dependence of the reaction rate. A mere increase from room temperature to 56.5 °C (the boiling point of acetone) can enhance the reaction rate by more than two orders of magnitude. This indicates that the catalytic reaction has a rather low barrier.

**Discussion.**

So far we have established that on pristine graphene surface, the C=O bonds in organics can be broken and polymers including polyethers can be formed. Meanwhile, the reaction products will not increase defect density in graphene and the interaction between graphene and the product molecules is mainly van der Waals. However, mild temperature annealing will not efficiently remove the polymer molecules because the molecules can be large. Therefore the polymers present a form of contamination on graphene and may affect the performance of devices thus made (Supplementary Fig. 3). So acetone is far from an ideal solvent to clean graphene. Still, the contamination can be totally removed by higher temperature annealing in UHV (Supplementary Fig. 2).

In summary, high-quality pristine graphene absent of heteroatoms was found to be catalytic active. It can react with organics including ketone, carboxylic acid and aldehyde which contain C=O bonds. The C=O bonds break and form chain-like and/or cyclic polymers under the catalysis of graphene as evidenced by Raman and STM measurments and DFT simulations. Inevitable defects on graphene can be the active sites, though other C sites cannot be excluded yet. The reaction products give rise to extra Raman peaks under 633 nm excitation. The product molecules mainly physically adsorb on graphene so they do not cause increase of defect density in graphene. The temperature dependence of these reactions are rather dramatic even at temperatures of ~50 °C. This discovery opens the door of using pristine graphene as a truly metal-free, dopant-free catalyst and should have important implications in polymer synthesis and graphene device fabrication.

**Acknowledgement:** The authors thank Prof. Kang L. Wang for helpful discussion. Z. Hu and X. Guo thank the National Key R&D Program of China (2016YFB0400603) for financial support. X. Guo thanks the National Natural Science Foundation of China (Grant 61335004).

**Author contributions:** Z.H. conceived the project. X.X. and C.L. conducted graphene growth and performed STM experiments. X.X. and F.K. performed Raman experiments. Z.L. and X.L. performed DFT calculations. All authors discussed the results, analysed the data and wrote the paper.

**Additional information:** Supplementary information is available in the online version of the paper. Reprints and permissions information is available online.

**Competing financial interests:** The authors declare no competing financial interests.

**Figure Captions:**

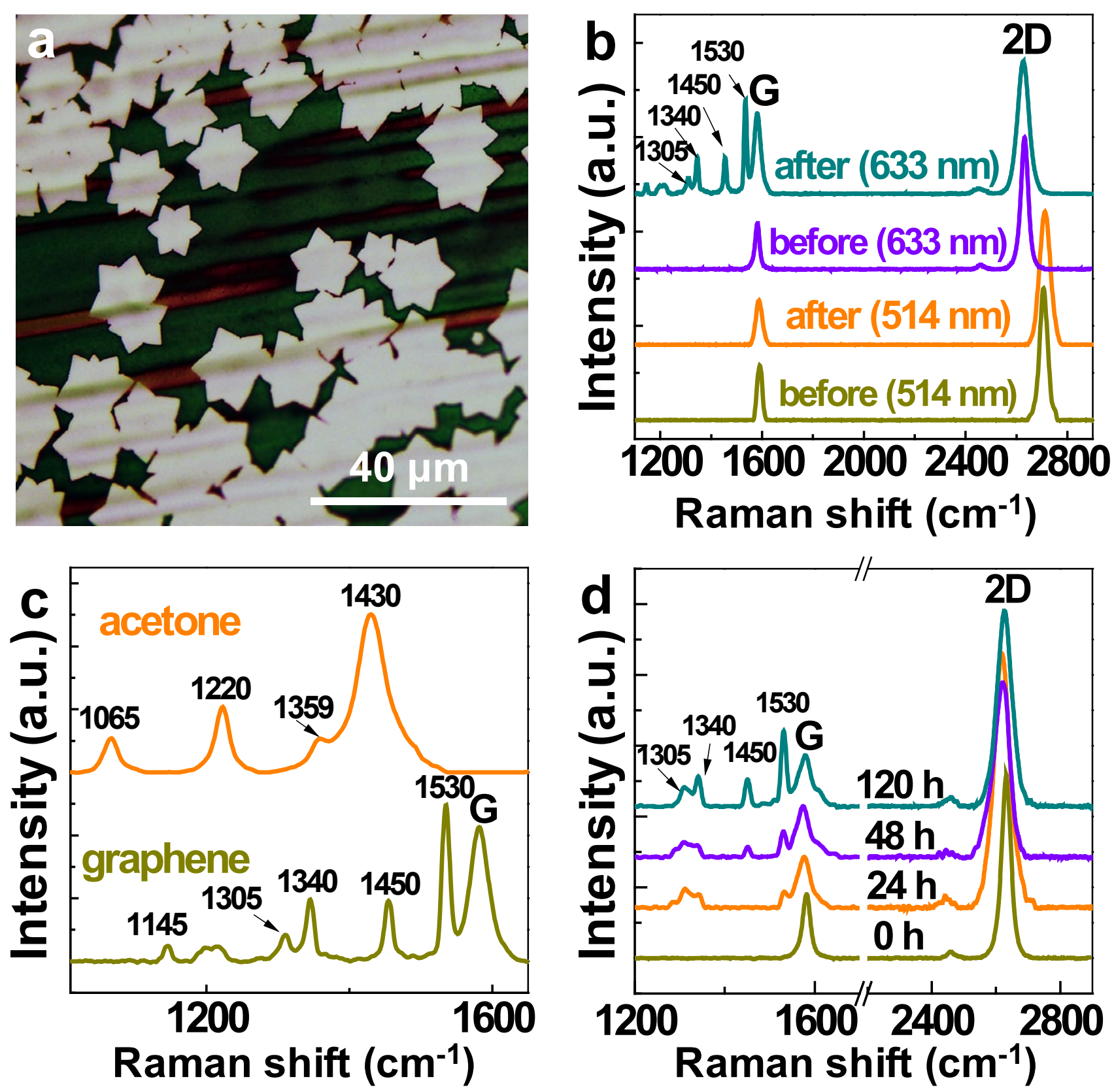

**Figure 1 | Extra Raman peaks after graphene immersed into acetone. a,** An optical image of a CVD grown graphene sample. The dark area is not covered by graphene and the snowflake-shaped bright features are single-crystal graphene grains. The scale bar is 40 μm. **b,** Raman spectra of graphene before and after being immersed in acetone. The excitation wavelenghths are 633 nm and 514 nm. The characteristic graphene G band and 2D band are present in all spectra. Extra peaks around 1305 $cm^{-1}$, 1340 $cm^{-1}$, 1450 $cm^{-1}$, 1530 $cm^{-1}$ are observed under 633 nm but not 514 nm excitation. **c,** Raman spectra (with 633 nm excitation) of pure acetone and graphene after acetone immersion, which indicate that the extra peaks are not from residual acetone. **d,** Raman spectra (with 633 nm excitation) of graphene samples immersed into acetone for 0 h, 24 h, 48 h and 120 h. The graphene G and 2D bands remain unchanged. The intensities of the extra peaks around 1305 $cm^{-1}$, 1340 $cm^{-1}$, 1450 $cm^{-1}$, 1530 $cm^{-1}$ increase with the immersion time. The Raman spectra in **b, c, d** are vertically displaced for clarity.

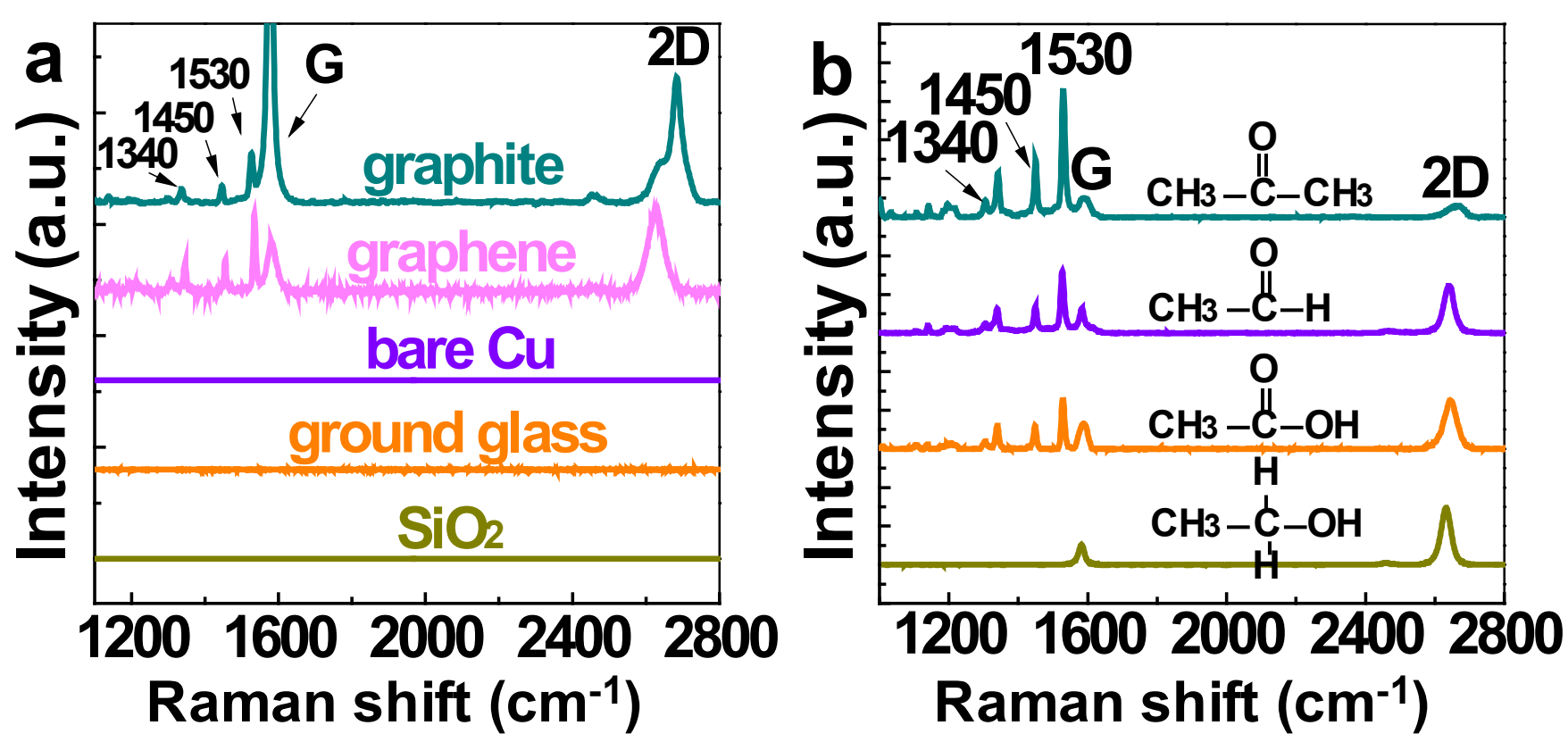

**Figure 2 | Check with various samples and organic solvents. a,** Raman spectra of various samples after being immersed into acetone. Only graphene and HOPG samples give the extra peaks. **b,** Raman spectra of graphene after being immersed in ethanol (brown), acetic acid (orange), ethanal (violet), and acetone (green). The same peaks around 1305 $cm^{-1}$, 1340 $cm^{-1}$, 1450 $cm^{-1}$, 1530 $cm^{-1}$ are observed in all samples but the one immersed in ethanol. These results point to the C=O bonds as the reaction bonds. The spectra are vertically displaced for clarity.

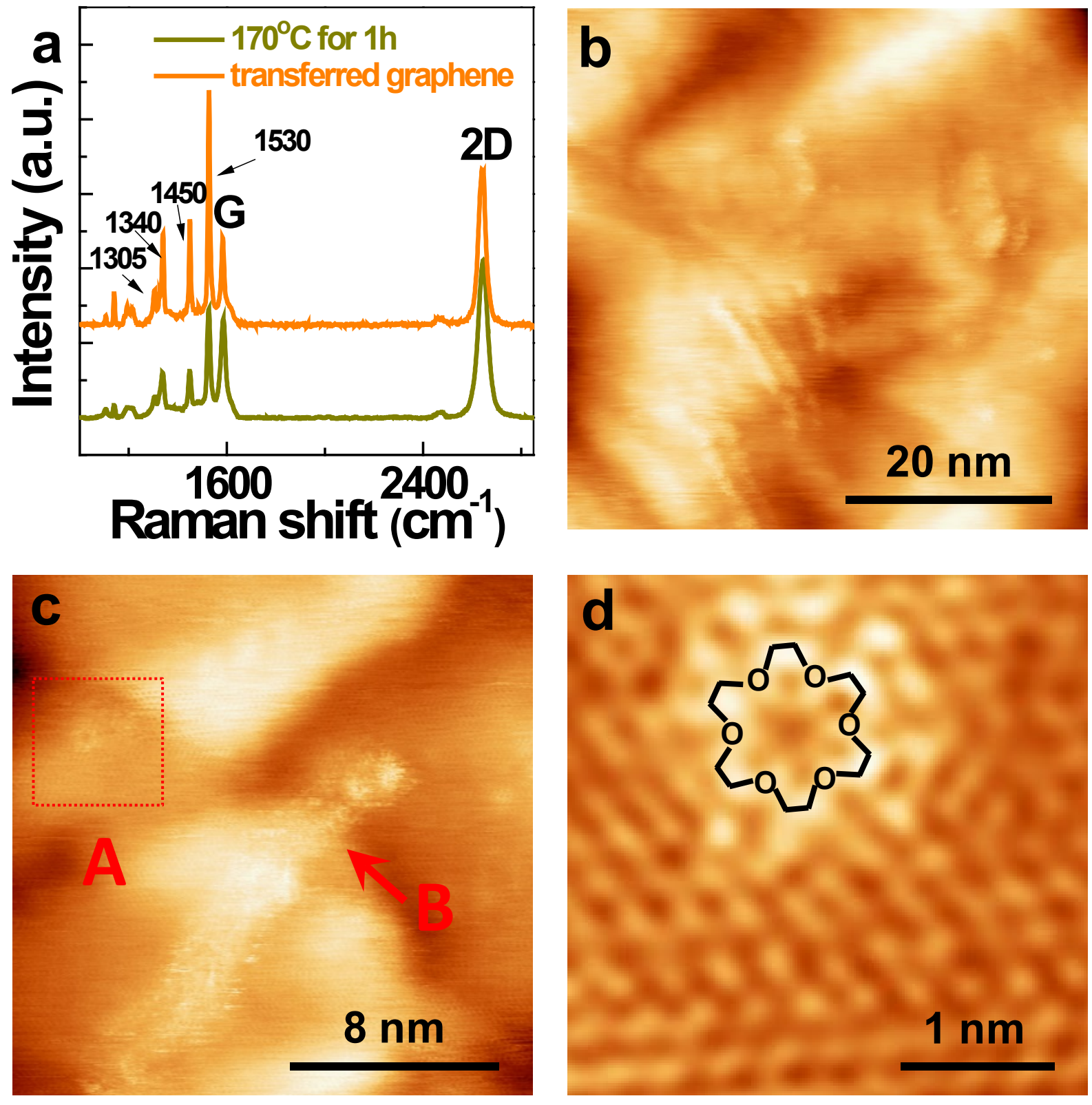

**Figure 3 | Investigation on the reaction products. a,** Raman spectra of samples before and after mild annealing at 170 °C in air. The extra peaks largely remain. The spectra are vertically displaced for clarity. **b,** A typical STM image of graphene after being immersed into acetone for 2 days. Lots of adsorbate protrusions (bright features) can be found. The scale bar is 20 nm. **c,** An STM image showing an area with adsorbates. There is a chain shaped feature marked by "B" and a ring shaped adsorbate marked by "A". The scale bar is 8 nm. **d,** Zoom-in image of the "A" region in **(b)**. The structure of the adsorbate takes the form of an 18-crown-6 crown ether. The scale bar is 1 nm.

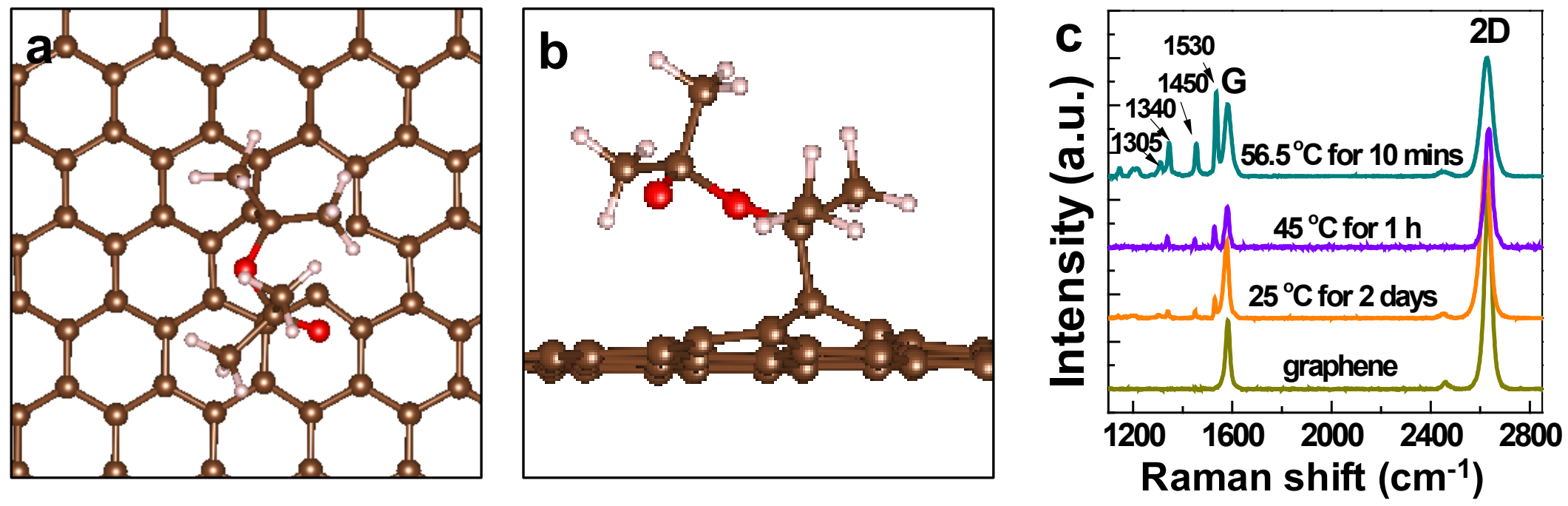

**Figure 4 | Investigation on the reaction energetics. a, DFT simulation results of the geometry optimized structure of graphene with defects and two acetone molecules,** top-view. **b,** side-view of the geometry optimized reacted structure. The C, O and H atoms are shown in brown, red, and white colors, respectively. **c,** Raman spectra showing the temperature dependence of the reaction rate. The brown curve corresponds to as grown graphene ($I_{2D}/I_G$ = 2.7, FWHM of 2D = 33 $cm^{-1}$). The intensities of the extra peaks (a reflection of the amount of the reaction products) increase dramatically with temperature.

## Methods

### Growth of graphene samples

We use atmospheric pressure chemical vapor depositon (APCVD) method to prepare the discontinuous graphene samples. Firstly, copper foil (25 μm, 99.8%, Alfa Aesar) was heated from room temperature to the growth temperature (1000 $^{o}$C) under the atmosphere of Ar (500 sccm) for 1h. Then, it was annealed for 40 minutes with $H_2$ (10 sccm) and Ar (500 sccm) flowing. After that, methane (5 sccm) was introduced as the carbon source. The growth process lasted for 2 minutes. Finally, we turned off the power and took it out of the tube furnace after it was naturally cooled to 200 $^{o}$C for about two and a half hours. Surface areas not covered by graphene would quickly oxidize in air and the color would change to red. Therefore we could readily observe graphene islands by optical microscopy (OM).

### Density functional theory calculation

The density-functional theory calculations were carried out using the VASP code, with the Perdew-Burke-Ernzerhof (PBE) functional for the description of the electronic exchange-correlation interactions. The projector augmented wave (PAW) potentials were used with a 300 eV plane wave cutoff energy. The total energies of free acetone and the defected graphene and the acetone-adsorbed graphene structure (labeled as acetone-graphene) were calculated employing a 5 × 5 graphene supercell with a 2 × 2 × 1 mesh of the Monkhorst-Pack k points.